\documentstyle[aasms4,flushrt]{article}

\begin{document}

\title{
THE GRAVITATIONAL LENSING IN THE QSO 1208+10 FROM THE PROXIMITY EFFECT
IN ITS LYMAN $\alpha$ FOREST}
\author{E. Giallongo$^1$, A. Fontana$^1$, S. Cristiani$^2$, S. D'Odorico$^3$
}

\bigskip

\affil{$^1$ Osservatorio Astronomico di Roma, via dell'Osservatorio, I-00040
Monteporzio, Italy\\
\noindent
$^2$ Dipartimento di Astronomia, Universit\`a di Padova, vicolo
dell'Osservatorio 5, I-35122, Padova, Italy\\
\noindent
$^3$ European Southern Observatory, Karl Schwarzschild Strasse 2,
D-85748 Garching, Germany\\
}
\begin{abstract}

The quasar Q1208+1011 ($z_{em}=3.8$) is the second highest redshift
double quasar ever detected. Several indications point toward it being
a gravitational lensed system, although a definitive proof is still
lacking.  We present new evidence of its lensed nature based on the
weakness of the ``proximity effect'' measured in the high resolution
Lyman absorption spectrum of the QSO. A luminosity amplification as
large as 22 has been derived from this analysis.  Indications on the
redshift of the lensing galaxy can be obtained from the analysis of
the intervening heavy element absorption systems discovered in the QSO
high resolution spectrum.  On statistical and dynamical grounds a MgII
system present at $z=1.13$ appears as the most likely candidate for
the lensing galaxy.  We compare the observed parameters with a simple
isothermal model for the lens to derive the properties of the lensing
galaxy.  The resulting magnification factor is smaller, although
marginally consistent with that derived by the analysis of the
proximity effect.

\end{abstract}


\section{INTRODUCTION}

High resolution images both from ground-based telescopes (Magain et
al. ,1992) and the HST (Bahcall et al. 1992a) revealed that the image
of the $z=3.8$ QSO Q1208+1011 (Hazard, McMahon \& Sargent 1986) is
split into two components, separated by 0.47'' and with a brightness
ratio of about 4, independent of wavelength.

Subsequent spectral observations from HST (Bahcall et al. 1992b)
confirmed that both images are from a $z_{em}=3.8$ QSO although small
differences in the emission line profiles were noticed.  Further
ground based imaging (Hjorth et al. 1995) has shown variability in the
intensity ratio of the continuum that scaled to 3.3 in the I band in
one year while a narrow-band filter centered on the Ly$\alpha$
emission showed again a more prominent Ly$\alpha$ in the brighter
source.

The observed variability and the failure in detecting the lensing
galaxy cast some doubt on the possibility that the double image
corresponds to a lensed source rather than a physical pair.  

In this work we show that additional evidence on the lensing nature of
this QSO can be obtained from the analysis of the proximity effect in
its Lyman $\alpha$ forest by means of new, high resolution spectra of
the QSO 1208+1011. The spectra of high--$z$ quasars at wavelengths shorter than
the Lyman $\alpha$ emission are dominated by the crowded sequence of
narrow absorption lines (the Lyman $\alpha$ forest), which are due to
a population of intergalactic clouds (Lyman $\alpha$ clouds)
intervening along the line--of--sight to the quasar. In addition,
absorption features due to other elements are more easily detected at
wavelengths longer than the Lyman $\alpha$ emission peak of the
spectrum. Systematic imaging and spectroscopic surveys have shown that
these heavy element (metal) systems are associated to intervening
galaxies at the same redshifts (e.g. Bergeron and Boisse 1991; Steidel
1995).

The proximity effect consists in a reduction of the Lyman $\alpha$
line density in the region near the QSO emission redshift due to the
increase of the ionizing flux by the QSO. It depends on the competing
ionization level produced by the diffuse UV background and the flux by
the nearby QSO. In this paper we show that the anomalously small
proximity effect measured in Q1208+1011 implies an intrinsic QSO
luminosity much smaller than observed suggesting a large amplification
of the QSO light due to the gravitational lensing effect.

The detection and above all the measurement of the redshift of the
lensing galaxy - presumably a faint object located in the middle of
the two point sources - is probably beyond current observational
possibilities.  For this reason, hints on its redshift have been
looked for in the absorption systems detected in the spectrum of
Q1208+1011.  Magain et al. (1992) presented a re-analysis of the
medium resolution spectrum of Q1208+1011 (Steidel 1990), identifying
19 possible metal systems, 14 of which in the range $2.5<z<3.1$.

In this paper we also discuss a MgII system at $z\simeq 1.13$ as
the most likely galaxy candidate responsible for the gravitational
lensing of the QSO. This identification has been proposed by Fontana
et al. (1997) and more recently by Semiginowska et al. (1998).

\section{THE LENSING MAGNIFICATION FROM THE PROXIMITY EFFECT}

The high resolution ($R=35000$) spectrum of Q1208+1011 was included in
the sample used by Giallongo et al. (1996) to estimate the UV
background at high redshifts from the proximity effect analysis. The
data were obtained at the ESO NTT telescope with the EMMI echelle
spectrograph in the framework of the ESO Key program 2-013-49
(P.I. S. D'Odorico). During the observations the slit was aligned
along the parallactic angle. Since the slit width (1 arcsec) and the
typical seeing (0.8 arcsec) were larger than the separation between
the two images of the QSO (0.47 arcsec), we assume to have collected
the spectra of both objects simultaneously. Four different exposures
were taken, for a total exposure time of $30000$s. The signal to noise
ratio of the final spectrum is greater than 20 per resolution element
in the regions of interest here.  We have analyzed all the absorption
features in the spectrum of Q1208+1011 deriving redshift, column
density and Doppler parameter of the absorbing cloud by fitting Voigt
profiles to isolated lines and individual components of the blends.
The detailed list of the absorption lines in this quasar will be given
in a separate paper. Here we discuss the implications that can be
inferred on the nature of the lensing system toward Q1208+1011.

The distribution of the Ly$\alpha$ absorption lines in QSO spectra
provides a powerful method to estimate the UV background (UVB) at high
redshift. Although their cosmological number density increases
strongly with redshift in the interval $z=1.5-5$ following a power-law
distribution $\propto (1+z)^{\gamma}$ with $\gamma \sim 2.7$, their
redshift distribution within a single QSO spectrum does not follow the
general cosmological trend when approaching the QSO emission
redshift. This effect has been interpreted as a ``proximity
effect''. It consists in a reduction of the line density in the region
near the QSO emission redshift due to the increase of the ionizing
flux by the QSO.Near the QSO the Ly$\alpha$ absorbers are more highly
ionized with a column density $N_{HI}=N_{\infty} / (1+\omega)$ where
$N_{\infty}$ is the intrinsic column density the same cloud would have
at infinite distance from the QSO and $\omega (z)= F/ 4 \pi J$ is the
ratio between the flux $F$ that the cloud receives from the QSO and
the flux $J$ that the cloud receives from the general UVB (see Bajtlik
et al. 1988 and Bechtold 1994 for details of the model).

The analysis of the proximity effect in the spectrum of the QSO
1208+10 shows that this QSO clearly stands out against the other
quasars of similar redshift because of the large number of saturated
absorption lines in the close vicinity of the quasar (see
Fig.~1). This suggests that the ionization produced by the quasar is
lower than for other QSOs of comparable apparent magnitude as expected
if Q1208+1011 is gravitationally amplified. To quantify this effect,
we have inverted the procedure used in Giallongo et al. 1996, i.e.  we
have fixed the ionizing UV background, the functional shape and the
parameters which best describe the statistical distributions of the
Lyman $\alpha$ clouds and obtained - through the same Maximum
Likelihood analysis - the {\it intrinsic} luminosity of this QSO.

Following Giallongo et al. (1996),  we have parameterized the 
cosmological distribution of the Lyman $\alpha$ clouds 
at any distance from the QSOs as:
\begin{equation}
{\partial ^2 n \over \partial z\partial N_{HI}}= A_o (1+z)^{\gamma} 
(1+\omega(z))^{1-\beta _f}
\end{equation}
$$
\times \left\{ \begin{array}{ll}
N_{HI}^{-\beta _f} & N_{HI} < N_{break}\\
N_{HI}^{-\beta _s} N_{break}^{\beta _s - \beta _f} 
& N_{HI} \geq N_{break} \end{array} \right.
$$
where the $(1+z)^{\gamma}$ term accounts for the redshift evolution of
the absorbers, the double power--law in $N_{HI}$ with slopes
$\beta_f$, $\beta_s$ accounts for the shape of their column density
distribution, and the $(1+\omega)^{1-\beta _f}$ term describes the
proximity effect.

To derive the intrinsic luminosity of Q1208+1011, we have computed a
Maximum Likelihood Analysis on the same line sample, fixing the
parameters describing the Ly$\alpha$ statistics and the value of $J$
to the result of Giallongo et al 1996, i.e. $\gamma = 2.65$, $\beta _f
= 1.35$, $\log N_{break} = 13.98$, $\beta _s = 1.8$, $J = 5
J_{-22}$ ($J_{-22} = 10 ^{-22}$ erg s$^{-1}$ cm$^{-2}$
Hz$^{-1}$sr$^{-1}$)
 and leaving the intrinsic Lyman Limit flux of Q1208+1011 as a
free parameter (see Tab. 3 in Giallongo et al. 1996 where the lines in
the region
affected by the proximity effect in Q1208+1011 were not used to minimize
the lensing bias).  The resulting flux predicted from this ``proximity
effect estimate'' is $F_{PE}=1.5 \times 10^{-28\pm 0.35}$ erg s$^{-1}$
cm$^{-2}$ Hz$^{-1}$, to be compared with the observed value
$F_{obs}=3.33 \times 10^{-27}$ erg s$^{-1}$ cm$^{-2}$ Hz$^{-1}$
estimated by Bechtold (1994).  As a further test, we have released the
previously mentioned parameters and performed a global fit obtaining
again the same QSO intrinsic flux as above and the best fit values as
in Giallongo et al. (1996).
The total amplification factor derived from this
analysis is $A = F_{obs} /  F_{PE} \simeq 22$. 

As discussed in the following session, current estimates of $J$
range from 5 to 10~$J_{-22}$ depending on the sample
and the statistical representation adopted. 
Since the estimated intrinsic QSO flux scales with the adopted
value  of the UVB, for
$J=10J_{-22}$ the derived intrinsic QSO flux would be 2 times higher,
leading to a correspondingly lower amplification. 

It is clear that this estimate is essentially statistical in nature.
Barring systematic effects - that will be discussed in the next
session - the major source of statistical uncertainity is the variance
of the Ly$\alpha$ lines statistical distribution.  We have estimated
it using a standard $\chi^2$ error analysis on the Maximum Likelihood
parameters, obtaining a $2\sigma$ upper limit $F_{PE}=7.5 \times
10^{-28}$, corresponding to a $2\sigma$ lower limit on the
amplification $A>4.4$ when a value of $J= 5 J_{-22}$ is used.

We therefore conclude that this result provides further support 
both to the gravitational lens hypothesis and to the 
photoionization model for the proximity effect, where it is expected to be
weaker in low luminosity QSOs.
As an alternative hypothesis, the
absorption systems in the environment of Q1208+1011 could be more gas-rich
than on average: in this case however we would have expected the
detection of the corresponding metal lines as well.

\section{DISCUSSION}

The reliability of the large amplification factor derived in Sect. 2
from the proximity effect analysis in Q1208+1011 depends on how the
estimates of both the UV background and the ionizing flux of the QSO
1208+1011 are unbiased.

Recent estimates of the UVB at $z\sim 3.5$ derived from the proximity
effect range from the value used in the present paper $J_{-22}=5\pm 1$
(or $5_{-1}^{+2.5}$ if we allow for asymmetric errors) to $J_{-22}\sim
10$ derived by Cooke et al. (1997). In fact the UVB estimate from
the proximity effect analysis depends on the slope of the column density
distribution of the weaker lines.  Giallongo et al. (1996) used the
flat slope $\sim 1.4$ derived from their data. This value is in good
agreement with the recent Keck results (Kim et al. 1997). The analysis
by Cooke et al. (1997) based on lower resolution and/or lower s/n data
assumes a single steep power-law distribution with a slope $\sim 1.7$
and results in a higher $J$ value.

Thus, an unbiased $J$ value definitely lower than $10^{-21}$ is found
(Giallongo et al. 1997), in agreement with what expected from the QSO
contribution at $z\sim 2-3$ (Haardt \& Madau 1996). There is also no
indication of any appreciable redshift evolution of the UVB in the
redshift interval $z=1.7-4.1$.

Gravitational lensing can also bias the estimate of the ionizing
UVB. If lensing brightens the QSO continuum then the QSOs are
intrinsically fainter than they appear, and $J$ is overestimated.
However, the fact that the deficiency of lines in QSOs of different
redshifts is correlated with luminosity but not with redshift suggests
that the statistical weight of the gravitational lensing effect is
small (Bechtold 1994).

The estimate of the ionizing flux of the QSO 1208+1011 can be affected
by variability (Hjorth et al. 1995). This implies that an accurate
measure of the amplification factor can only be obtained by monitoring
the luminosities of various images on a time interval of at least a few
years.

In summary, the unbiased statistical estimate of the UVB at $z\sim
3.5$ together with its image splitting in two components strongly
support the gravitational lensing hypothesis for Q1208+1011.

Since the very limited spatial separation of the two components
prevents the detection of the lensing galaxy, hints on its redshift
have been sought from the heavy element systems in the QSO spectrum.
Early investigations at low resolution (Steidel 1990) revealed the
existence of an unusually crowded series of CIV absorption systems at
$z\simeq 2.8 - 2.9$.  The identification of one of this high $z$
absorption system with the lensing galaxy is not straightforward.
Indeed, the small angular separation of the lens corresponds to an
impact parameter of $\sim 2$ kpc at $z=2.9$.

However, CIV systems are expected to arise preferentially at large
impact parameters, while the very small impact parameters implied for
this lens should favor the detection of low--ionization species.
Besides the cluster of absorption systems at $z\geq 2.8$, an
interesting candidate for the lensing galaxy is the $z=1.13$ MgII
absorption system (Fig. 2) that we discovered from the analysis of our
high resolution spectrum (Fontana et al. 1997, see also Siemiginowska
et al. 1998).  As can be seen from fig.2, the identification doesn't
rely simply on the wavelength coincidence, but on the good agreement
in the relative intensities of the lines.  This MgII complex has been
fitted with six components, which is the minimum number required to
give a satisfactory fit.

The total equivalent width is $\simeq 0.7$ {\AA}.  Several systems
with this equivalent width have been detected in MgII surveys at very
small impact parameter (e.g. Steidel 1995).

Several arguments indicate that the $z=1.13$ MgII absorption
system may be connected with the lensing galaxy.  As shown by
Kochaneck (1992), statistical arguments favour intermediate redshifts
for the lensing galaxy. The probability distribution of the redshift
of the lensing galaxy is a strongly peaked function, that in the case
of Q1208+1011 has a maximum at $z=1-1.15$ depending on cosmology.  At
larger redshift, this probability becomes rapidly smaller, being
about $0.008$ at $z\sim2.8$. Great care has to be taken in applying
this statistical approach to a single object, to avoid the pitfalls of
``a posteriori statistics''. We note however that these models have a
direct physical ground: lenses are most efficient when placed at half
way of the light path.

In a simple isothermal model for the lensing galaxy, the 1--D velocity
dispersion can be computed from the image separation $\Delta \theta$
alone, as (Turner, Ostriker and Gott 1984): $\sigma_{\parallel} = c
(\Delta \theta S /(8 \pi))^{1/2}$ where S accounts for the cosmology
and is weakly dependent on $\Omega$.  At $z=2.9$ we derive a 1--D
dispersion $\sigma_{\parallel} \sim 300$ km s$^{-1}$ to be compared
with an observed value in the CIV absorption systems of about 240 km
s$^{-1}$. The corresponding dynamical mass enclosed in the impact
parameter $\sim \Delta \theta /2 \simeq 1.75$ kpc is $M\sim 10^{11}$
M$_{\odot}$ and the corresponding virial mass is $M_L \sim 3\times
10^{12}$ M$_{\odot}$. These values are hardly credible at $z\simeq
2.9$ in any cosmological scenario for galaxy formation. The $z=2.9$
CIV systems could provide a mass overdensity (extended over several
hundreds of kpc) not related to the split of the QSO images but able
to provide a further luminosity amplification at the most.

Alternatively, the velocity dispersion expected at $z=1.13$ is
$\sigma_{\parallel} = 140$ km s$^{-1}$, the same observed in the MgII
system (Fig.~2) with an enclosed mass in a radius $\sim 2$ kpc of
$M\sim 3\times 10^{10}$ M$_{\odot}$ and a virial mass $M_L\sim 7\times
10^{11}$ M$_{\odot}$.

The comparison between predicted and observed velocity dispersion
assumes that the observed gas velocity spread is representative of the
galaxy velocity dispersion. Although there are hints favouring this
association from MgII absorptions detected both in the halos of
external galaxies and in our own galaxy (e.g. Lanzetta and Bowen 1992;
Bowen, Blades, \& Pettini 1995), there are also examples of strong
absorptions spanning $\sim 300$ km s$^{-1}$ which seem unrelated to
the velocity dispersion within galaxies (Bowen, Blades, \& Pettini
1996).  

In conclusion, the $z=1.13$ galaxy giving rise to the MgII
absorption system appears as the best candidate for being the lensing
galaxy.
In the same simple isothermal model used above, the total
amplification $A$ is a simple function of the observed brightness
ratio between the two components $R$: $A = 2 \times (R+1)/(R-1)$
yielding $A\simeq 3 (4)$ if $R\sim 4 (3.3)$, as found in the two
available measurements of Q1208+1011, i.e. a value which is near the
$2 \sigma$ lower limit derived from the analysis of
the proximity effect, leaving the possibility that an additional
magnification mechanism exists.

Finally, we would like to point out two opportunities for future work.

Q1208+1011 and its surrounding field deserve  deeper
investigations, to search for the lensing galaxy and  other foreground
structures. The HST or a ground-based telescope equipped with
adaptive optics could be used to provide evidence for the faint
lensing galaxy at $z\sim 1$. The observations should be carried at
wavelengths shorter than 580 nm, where the continuum of the QSO is
depressed by the IGM absorption. 

There are 7 lensed QSOs known to date at $z>2.5$ that are within the
capabilities of high resolution spectrographs at very large
telescopes: they could be used to confirm on a statistical basis the
validity of this approach and to provide further constraints on the
lensing mechanism.

\acknowledgments 
We are grateful to the referee A. Songaila for suggestions which
improve the clarity of the paper. We are also grateful to M. Vietri
for stimulating discussions.

\newpage

\centerline{\bf FIGURE CAPTIONS}

\noindent
Fig.~1. The Ly$\alpha$ forest affected by the proximity effect in two
QSOs of comparable apparent magnitude $V\sim 17.5$ and at the same
$z\sim 3.8$. Upper panel Q2000-33, lower panel Q1208+1011.

\noindent
Fig.~2. The MgII doublet at $z=1.13$. Six components have been fitted.

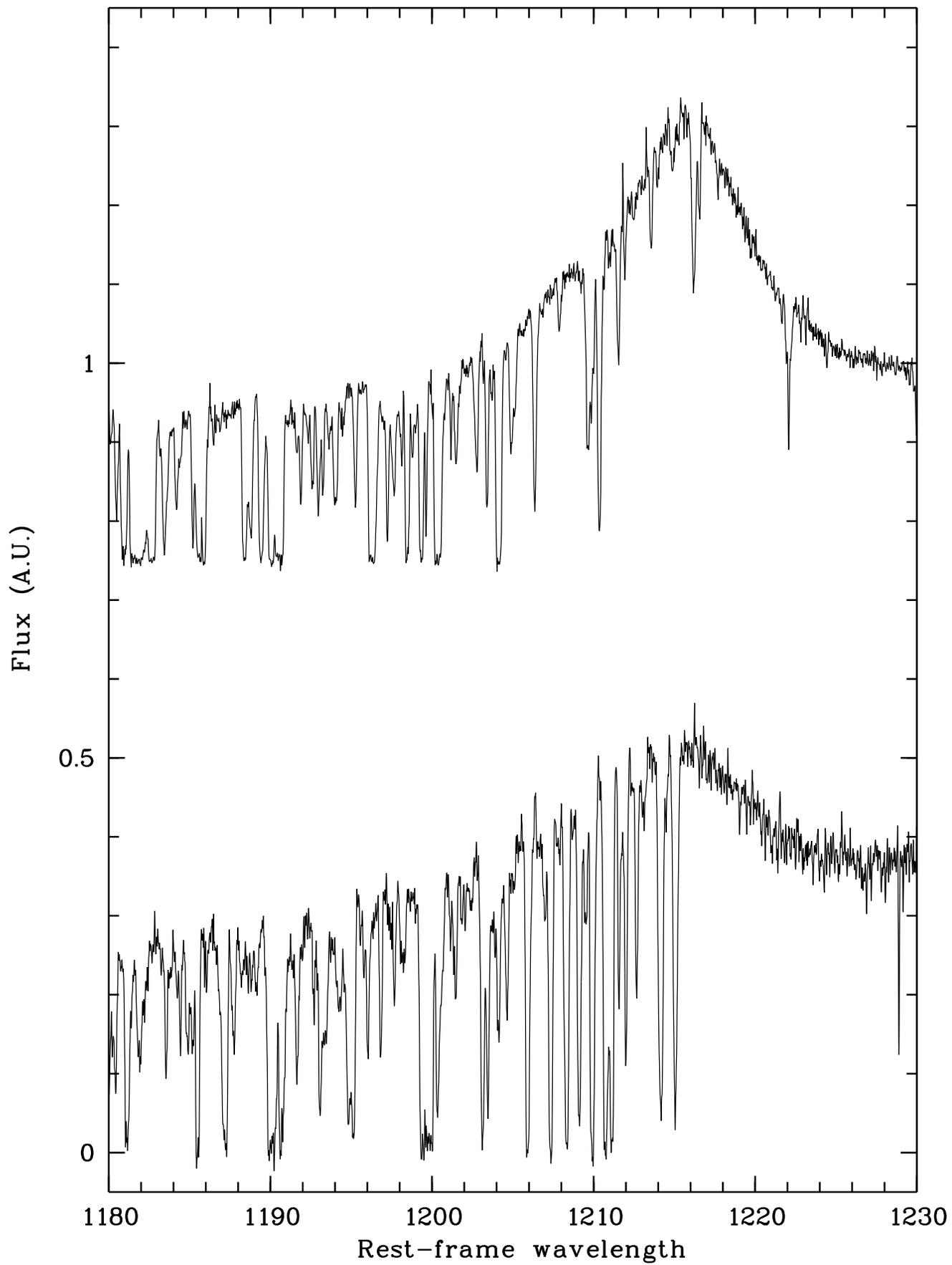
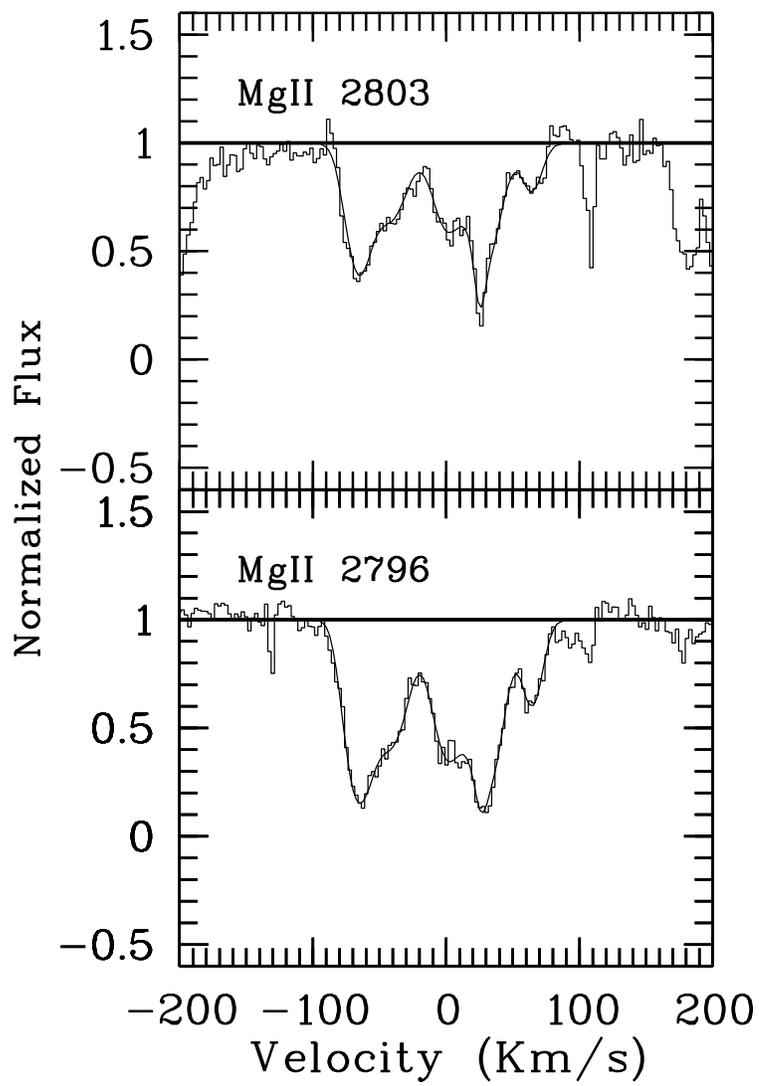
\end{document}